# Metamaterials with Negative Compressibility Highlight Evolving Interpretations and Opportunities

Zachary G. Nicolaou[1], Feng Jiang[2,3] & Adilson E. Motter[2,3,4,5]

[1] Department of Applied Mathematics, University of Washington, Seattle, WA 98195, USA
[2] Department of Physics and Astronomy, Northwestern University, Evanston, IL 60208, USA
[3] Center for Network Dynamics, Northwestern University, Evanston, IL 60208, USA
[4] Department of Engineering Sciences and Applied Mathematics, Northwestern University, Evanston, IL 60208, USA
[5] Northwestern Institute on Complex Systems, Northwestern University, Evanston, IL 60208, USA

**The idea that a material can exhibit negative compressibility is highly consequential for research and applications. As new forms for this effect are discovered, it is important to examine the range of possible mechanisms and ways to design them into mechanical metamaterials.**

Intuition, elementary physics, and formal thermodynamic stability arguments all suggest that a material under uniform compression must respond by reducing its volume. In elasticity theory, this response is characterized by the material's compressibility, defined as the relative reduction in volume $-dV/V$ divided by the increase in applied pressure $dP$ (i.e., the reciprocal of the bulk modulus) [1]. Per this definition, the compressibility of a material in thermodynamic equilibrium is indeed strictly positive.

However, physical phenomena are diverse, and there are important scenarios in which this no-go theorem can be circumvented by relaxing one or more of the underlying assumptions, namely: (i) that the process is continuous, (ii) that there is no net macroscopic exchange of mass or energy with the environment, and (iii) that the system is not subject to macroscopic changes. New (and often surprising) possibilities emerge by departing from continuous, closed, and time-independent conditions, as illustrated by Caprini *et al.* [2] for a system that expands discontinuously as it undergoes fluid intrusion in response to an increase in hydrostatic pressure.

**Mechanisms for negative compressibility**

Several mechanisms can be identified for understanding complementary notions of negative compressibility (see Fig. 1). These mechanisms can be explored in mechanical metamaterials, which are engineered to gain their properties from their structure rather than purely from their bulk composition, thereby allowing unconventional responses. Metamaterials have potentially transformative technological applications, such as in the development of protective devices [3], mechanical computing [4], and soft robotics [5].

While (static) negative compressibility cannot occur continuously under the standard assumptions above, metamaterials can be designed to undergo a negative compressibility transition in which they change discontinuously from a low- to a high-volume phase in response to increased pressure [6] (Fig. 1a). The effect is enabled by the stress-induced decay of metastable states, involves a finite increase in volume in response to a differential increase in applied pressure (also characterized in terms of finite-difference compressibility $-\frac{1}{V}\frac{\Delta V}{\Delta P}$), and can be cycled through a hysteresis loop. Negative compressibility transitions were proposed over a decade ago [6], but they have only recently been experimentally achieved with buckling designs constructed using additive manufacturing [7].

A different mechanism has been examined in open poroelastic materials, where pressure is applied via a fluid medium that penetrates the material [8] (Fig. 1b). When the material is fully penetrated, the medium pressure pushes out from within the material as much as it acts to compress it. However, the differing responses of subcomponents can still lead to negative effective compressibility (the "effective" volume here includes that occupied by the penetrating medium). Naturally, the compressibility of the combined (closed) system consisting of the penetrated material and the surrounding fluid medium remains positive (as does the compressibility of the bulk poroelastic material itself). This form of negative compressibility is especially relevant in small-scale stochastic systems, such as in the negative response exhibited by some proteins as they denature under increasing pressure [9].

A third possible mechanism involves sources of energy other than the mechanical work done by the applied pressure (Fig. 1c). Although this mechanism remains largely unexplored as a source of negative compressibility, we suggest that it has potential for future metamaterial design.

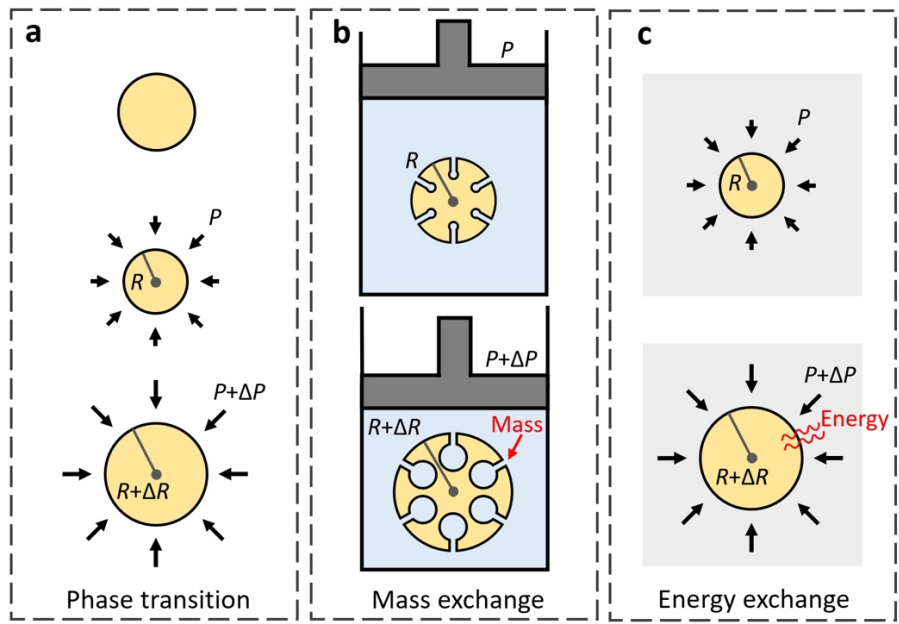

**Figure 1. Schematics of different mechanisms to achieve negative compressibility. a** Negative compressibility transition in a closed system. The material radius initially decreases as the applied pressure increases. The radius then suddenly increases when the pressure passes a critical threshold, and the material undergoes a negative compressibility transition. **b** Negative compressibility induced by mass exchange in an open system. When the piston is compressed, the hydrostatic pressure increases, and the fluid (blue) more fully penetrates the matrix of the poroelastic material (yellow). As a result of differing responses across the matrix, the effective volume of the material increases, giving rise to a form of negative compressibility. **c** Negative compressibility induced by energy exchange in an open system. As the material is stressed, it simultaneously absorbs energy from the surrounding environment

(grey). This energy is converted into mechanical work in the material, causing it to expand against the applied pressure and exhibit negative compressibility. In all panels, $R$ and $P$ indicate radius and applied pressure, respectively, and $\Delta R$ and $\Delta P$ indicate the corresponding increments.

**Metastable elastocapillary systems**

In their recent study [2], Caprini *et al*. propose a simple model system that combines transitions from metastable states with open poroelastic metamaterials, giving rise to a new kind of negative compressibility transition.

Unlike previous poroelastic systems, the immersed porous material (schematically represented by a cavity supported by springs in Fig. 2) is hydrophobic, leading to the energetically favorable formation of a vapor bubble that fills the cavity at low pressures. As long as the fluid is excluded by the vapor, the cavity represents a closed system, and the volume decreases with increasing pressure (Fig. 2a-b).  However, the energetics reverse for sufficiently large pressure, leaving the system in a metastable state until this state decays, resulting in fluid intrusion into the cavity. Once intrusion occurs, the fluid pressure no longer compresses the cavity, acting equally within and outside it, and the springs relax (Fig. 2c). This causes the system to expand through a (discontinuous) negative compressibility transition (Fig. 2b-c). Cycling the pressure leads to repeated hysteretic negative compressibility transitions as the metastable bubbles vaporize and collapse (Fig. 2d-e).

The spring-fluid model is appealingly simple, relying only on basic thermodynamic assumptions, and it applies across many scales, as experimentally demonstrated for nano- and millimeter-scale systems. The results represent an important paradigm that will aid future research on negative compressibility.

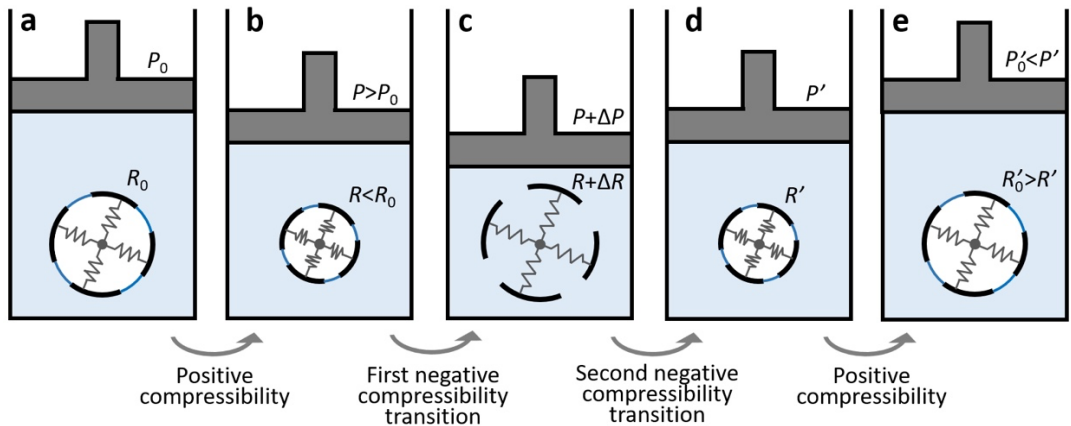

**Figure 2. Cycle of intrusion and extrusion in a spring-fluid metastable model exhibiting negative compressibility.** The black dashed line marks the boundary of a hydrophobic cavity, the blue (white) background represents the water (vapor), and the dark blue line indicates the gas-liquid interface. **a-b** Positive compressibility in a vapor-filled cavity, where the cavity radius is determined by the balance between the applied hydrostatic pressure and the internal restorative force (represented by the springs). The radius decreases continuously as the pressure is progressively increased. **c** Negative compressibility transition as fluid intrusion occurs, where the cavity expands as the pressure passes a critical threshold. **d** A second negative compressibility transition as fluid extrusion occurs upon pressure reduction. **e** End of the cycle following a second leg of positive compressibility, in which the radius increases upon pressure reduction. Intrusion and extrusion cycles allow the system to switch

between states, leading to repeatable negative compressibility transitions. The mathematical symbols indicate radii and pressures, mirroring the notation in Fig. 1. (Image developed based on Ref. [2].)

**Seemingly related properties**

Negative compressibility, as discussed above, is a longitudinal effect: the volume, area, or length deformation is measured along the dimensions of the applied force. An interesting but very different class of negative material properties concerns responses in directions transverse to an applied force. Examples include auxetics with negative Poisson's ratio [10] and stretch-densifying materials [11]. They also include so-called negative linear or area compressibility materials, which exhibit a negative response in one or two dimensions that is compensated by a positive response in the remaining dimension(s) [12,13]. Materials with these properties have found interesting applications, but they do not generally exhibit negative volumetric compressibility.

Some studies have also referred to negative stiffness inclusion to describe composite materials in which an increase in deformation leads to a decrease in the inclusion's restoration force [14]. This corresponds to a negative slope branch in the force-deformation diagram, which can be occupied when the control parameter is the deformation. Crucially, the notion of compressibility requires the control parameter to be the force, and negative slope branches (corresponding to negative compressibility) are unstable in force-controlled experiments under the standard assumptions (i)-(iii) above. Therefore, under these assumptions, experiments can only be used to measure compressibility along branches of positive compressibility.

Notions of elastic moduli have also been generalized to dynamical quantities. For a constituent of a phononic material, stiffness can be defined as the ratio $F(t)/x(t)$ between the applied time-dependent force $F(t)$ and the resulting displacement $x(t)$. Negative dynamical moduli are thus possible for the material when the driving and the response are out of phase with one another [15]. Compelling applications, such as acoustic cloaking and superlensing [16], are now well-established in such metamaterials, but we stress that these negative quantities are conspicuously different from the (static) notions of negative compressibility discussed above.

**Opportunities for future research**

Given the roles of discontinuity and poroelasticity in previously identified negative compressibility and the importance of metamaterial design for discovering new responses, what other mechanisms can give rise to genuine forms of the phenomenon, and what other guiding principles lie ahead for the field? One promising future direction is to explore mechanisms for exchanging energy with the environment, as schematically illustrated in Fig. 1c. Because materials can be modeled as mechanical networks, employing tools from network science [17,18] could provide valuable insights for designing and analyzing new metamaterials. A limited application of this approach has been successfully employed to design constituents for negative compressibility transitions [6], but its potential for network physical learning and the study of other systems and compressibility responses remains underexplored. Network approaches may, in particular, facilitate the design of active matter systems, where energy-driven non-equilibrium processes give rise to a plethora of interesting phenomena [19,20]. The study of compressibility in active matter systems is a promising uncharted frontier.

**Acknowledgements**

The authors acknowledge support from ARO MURI grant No. W911NF-22-2-0109.

**Authors and Affiliations**

Zachary G. Nicolaou[1], Feng Jiang[2,3] & Adilson E. Motter[2,3,4,5]

[1] Department of Applied Mathematics, University of Washington, Seattle, WA 98195, USA
[2] Department of Physics and Astronomy, Northwestern University, Evanston, IL 60208, USA
[3] Center for Network Dynamics, Northwestern University, Evanston, IL 60208, USA
[4] Department of Engineering Sciences and Applied Mathematics, Northwestern University, Evanston, IL 60208, USA
[5] Northwestern Institute on Complex Systems, Northwestern University, Evanston, IL 60208, USA

*All authors contributed equally to this work.*

**Corresponding author**

Correspondence to Adilson E. Motter (motter@northwestern.edu).